# Polymeric Frameworks as Organic Semiconductors with Controlled Electronic Properties

*Ken Sakaushi,*[†,§,#,*] *Georg Nickerl,*[§] *Hem Chandra Kandpal,*[‡,+] *Laura Cano-Cortés,*[‡]

*Thomas Gemming,*[†] *Jürgen Eckert,*[†,¶] *Stefan Kaskel,*[§] *and Jeroen van den Brink*[‡,*]

[†]IFW Dresden, Institute for Complex Materials, Helmholtzstr. 20, D-01069 Dresden, Germany

[§]TU Dresden, Department of Inorganic Chemistry, Bergstr. 66, D-01069 Dresden, Germany

[+]Indian Institute of Technology Roorkee, Department of Chemistry, Roorkee-247 667, India

[¶]TU Dresden, Institute of Materials Science, Helmholtzstr. 7, D-01069 Dresden, Germany

[‡]IFW Dresden, Institute for Theoretical Solid State Physics, Helmholtzstr. 20, D-01069 Dresden, Germany

AUTHOR INFORMATION

**Corresponding Author**

* ken.sakaushi@mpikg.mpg.de (K.S.); j.van.den.brink@ifw-dresden.de (J.v.d.B.)

Present Addresses

[#]Max Planck Institute for Colloids and Interfaces, Research Campus Potsdam-Golm, D-14424 Potsdam, Germany.

**ABSTRACT.** The rational assembly of monomers, in principle, enables the design of a specific periodicity of polymeric frameworks, leading to a tailored set of electronic structure properties in these solid-state materials. The further development of these emerging systems requires a combination of both experimental and theoretical studies. Here, we investigated the electronic structures of two-dimensional polymeric frameworks based on triazine and benzene rings, by means of electrochemical techniques. The experimental density of states was obtained from quasi-open-circuit voltage measurements through galvanostatic intermittent titration technique, which we show to be in excellent agreement with first principles calculations performed for two and three-dimensional structures of these polymeric frameworks. These findings suggest that the electronic properties do not only depend on the number of stacked layers but also on the ratio of the different aromatic rings.

**TOC GRAPHICS**

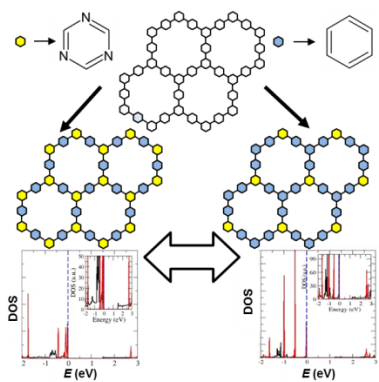

**KEYWORDS**. Electrochemistry; Electronic property; Organic semiconductors; Polymeric Frameworks; Rational assembly.

Since the discovery of conductive polymers[1–3], semiconducting conjugated polymers have been of great interest in organic electronics applications[4,5], such as organic electro-luminescent diodes[6–9], photovoltaic-cells[10,11], photocatalyst polymers[12] and organic batteries[13–16]. One of the most interesting properties pursued by organic electronics is bipolarity: existence of both *p*- and *n*-type semiconducting behavior in the same material[17]. This feature is not only important from fundamental science but also for its possible applications: bipolar organic compounds are promising candidates to promote further development of the field of organic electronics[18]. The discovery of bipolarity from a new class of organic material would foster future developments of the above mentioned technologies[19]. Indeed, recent works on π-conjugated microporous polymers show that polymeric frameworks are promising materials for organic electronics[20,21] and even for organic spintronics[22]. The experimental control of structural periodicities by choosing a monomer as the building block of the framework can lead to semiconducting systems with unique electronic properties, such as two-dimensional (2D) atomic crystals[23]. The possibility of controlling bipolar organic semiconductors electronic properties can give rise to new electronic system-level design[24] based on artificial semiconducting polymeric frameworks, which would represent a giant leap forward in the development of organic electronic devices.

Here, we studied the electronic properties of covalent triazine-based frameworks[25] (CTFs) by performing electrochemical measurements and comparing with first principles electronic structure calculations. The CTFs are porous polymeric frameworks formed by cyclotrimerization of nitrile monomers, which have been applied in the implementation of catalyst materials[26,27], and most recently for lithium- and sodium-based energy storage devices[28,29]. They have a conjugated structure, consisting of benzene rings as electron donors and triazine rings as electron acceptors (Fig. 1), with controllable photoluminescent properties[30]. Despite the above mentioned properties, surprisingly, little work has been done in the study of CTFs towards an efficient implementation in organic electronics. In a previous study, we have carried out several electrochemical measurements to test the electrochemical properties of electrode materials, such as CTFs[28,29]. From these experiments, important information about the electronic structure of materials[31–33] can be obtained. Thus, we applied these electrochemical techniques in the present work to investigate the tunability of the electronic properties of porous polymeric frameworks, testing their properties as organic semiconductors, which we combined with a theoretical study of the density of states for two CTFs compounds.

We synthesized a CTF porous polymeric framework consisting of benzene and triazine rings in the theoretical ratio of 1:1.5, so called CTF-1. By performing a careful tuning of the synthetic conditions, we can obtain the crystalline form of CTF-1, which has a 2D structure with an eclipsed AAA stacking, and *P6/mmm* symmetry. We characterized CTF-1 by using spectroscopic measurements. In this sense, Raman spectroscopy measurements (RSMs) can reveal the electronic structure of materials, as well as information about the associated structural periodicity. Since the structures of the analogous organic systems graphite and graphene have been extensively studied by RSMs[34–38], we compared the Raman spectra of CTF-1 and graphite (Fig. 2a). In this case, the multilayer and monolayer CTF-1 is comparable to N-doped porous graphite and N-doped porous graphene, respectively. The *G* peak is produced by the doubly degenerated zone center $E_{2g}$ mode which corresponds to the motion of the atoms in the 2D honeycomb structure[34-36]. The *D* peak, which is an inactive mode for perfect graphite, is induced by disorder in the 2D structure[34,36]. The peak at ~2700 cm$^{-1}$ is historically called *G'* peak, but using the same notation as in refs. 36 and 38, we call this peak *2D*. The comparison of the *D*-to-*G* intensity ratio[37], *I(D)/I(G)*, for CTF-1 (1.1) and graphite (0.3), suggests that CTF-1 has a shorter 2D ordered structure (Fig. 2a). However, the existence of the *G* peak for CTF-1 clearly reveals the formation of a 2D honeycomb structure constituted by benzene and triazine rings. Electron energy-loss spectroscopy (EELS) measurements were carried out to have further information about the structure of CTF-1 (Fig. 2b). The fine features of C-*K* and N-*K* edges indicate a *sp*$^2$-bonding, characteristic of graphitic networks[39]. The 1s → π* transition observed confirms the *sp*$^2$-hybridization for carbon and nitrogen atoms at ~285 eV (Fig. 2b) and ~400 eV (Fig. 2b inset), respectively. Therefore, by performing Raman and EELS experiments, we confirmed the formation of a 2D polymeric framework constituted by aromatic rings. In addition, high resolution transmission electron microscopy (HR-TEM) observation (Fig. 2c), showed the stacking of 2D sheets of CTF-1 with ~10 layers (Fig. 2c inset). The porous structure of CTF-1 cannot be clearly observed due to multiple sheets stacking and the resolution of HR-TEM imaging. However, N$_2$ physisorption experiments (Fig. 2d) indicate the existence of a porous structure of the CTF-1 with a pore-diameter of ~14 Å, which is in good agreement with the pore-size of the perfect crystalline CTF-1 (Fig. 2d). These results reveal the formation of CTF-1 with 2D ordering. Therefore, covalent triazine polymeric frameworks show controllable electronic properties and semiconducting character due to a graphene-like structural periodicity, combined with a suitable choice of the monomer as building block.

We measured open-circuit voltage (OCV) (=$E_{OCV}$) curves through galvanostatic intermittent titration technique (GITT) to investigate the quasi-equilibrium states[40,41] of CTF-1 (Fig. 3a,b). The bipolarity observed in this material is derived from the electron-donor and electron-acceptor character of the benzene and triazine rings, respectively[25,28]. Thus, CTF-1 can have both an *n*-doped state (negatively charged state; Fig. 3c) and a *p*-doped state (positively charged state; Fig. 3d). It has been suggested in previous electrochemical measurements on the porous polymeric frameworks as electrodes with LiPF$_6$ in ethylene carbonate/dimethyl carbonate (1:1) as the electrolyte that Li$^+$ and PF$_6^-$ are coordinated with CTFs based on a redox mechanism[28]. Electrochemical properties of polymeric frameworks as electrodes are directly connected with their electronic structures[42,43]. For the *n*-doping process (Li$^+$ reaction), we found a plateau at ~2.7 V vs. Li/Li$^+$ and following sloping curve in a discharge curve (Fig. 3a). The slope of this OCV curve is different from the ones obtained for typical intercalation compounds, which show a plateau followed by a sharp drop in the potential[32]. In the *p*-doping process (PF$_6^-$ reaction) we observed a plateau at ~4.2 V vs. Li/Li$^+$ (Fig. 3b). Although the details of the energy storage mechanism of porous polymeric frameworks are still unclear, previous works on the adatom of both cations and anions into graphite[44–46], graphene[45–49], and related materials suggest that the ions could be coordinated on the top of the aromatic rings, which serve as hosts. This unique electrochemical reaction should correspond to the electronic structure of CTF-1.

Based on the above discussion, we described the relation between the OCV measurements and the electronic structure of CTF-1, by means of thermodynamics and statistical mechanics[31–33,50]. The potential given by OCV depends on the chemical potential, $\mu$, of the guest (Li$^+$), and it is described by Nernst equation as follows:

$$E_{OCV} = \frac{-\mu}{e} \quad . \tag{1}$$

We considered the lattice-gas model, in which each site of the lattice has two states: full or empty. Then, we can define the occupancy of the guest sites, $f(\varepsilon)$, as a function of energy, $\varepsilon$, through the Fermi distribution:

$$f(\varepsilon) = \frac{n(\varepsilon)}{N(\varepsilon)} = \frac{1}{1+\exp\left(\frac{\varepsilon-\mu}{k_B T}\right)} \quad , \tag{2}$$

where $N(\varepsilon)$ is the number of sites with energy $\varepsilon$, and $n(\varepsilon)$ is the number of occupying Li ions. From Eq. 1 and Eq. 2, we derive the relation between $n(\varepsilon)$ and $E_{OCV}$:

$$n(\varepsilon) = \frac{N(\varepsilon)}{1+\exp\left(\dfrac{\varepsilon + eE_{OCV}}{k_B T}\right)} \quad . \tag{3}$$

Eq. 4 shows the integration of $n(\varepsilon)$ over the entire energy range, giving us the number of intercalated Li ions in the polymeric framework, $n$:

$$n = \int_{-\infty}^{+\infty} \frac{g(\varepsilon)}{1+\exp\left(\dfrac{\varepsilon + eE_{OCV}}{k_B T}\right)} d\varepsilon \quad , \tag{4}$$

here we have defined the distribution of the site energy as $g(\varepsilon) = dN(\varepsilon)/d\varepsilon$.

For ideal bulk electrodes, one finds a flat plateau in the OCV potential curves, due to the fact that all sites have a specific energy, $\varepsilon_i$, thus corresponding $g(\varepsilon)$ to a delta function $\delta(\varepsilon - \varepsilon_i)$ in Eq. 4. In the case of anisotropic and/or disordered electrode materials, the value for the occupation energy is broadened: $\Delta\varepsilon$. This energy distribution for the sites of the guest is reflected on the change in the slope of OCV curves, where the steepness that is observed for ideal electrodes is reduced, resulting in smoother slopes. Thus, if we assume a monolayer or completely disordered CTF-1, following Eq. 4, we would expect a smooth slope OCV curve without any plateau. This can be understood from the modification of the electronic structure of CTF-1, due to both anisotropy and the existence of defects[31–33]. However, CTF-1 shows a plateau, which confirms that this polymeric framework is partially formed by an ordered multilayer structure (Fig. 2c).

The distribution of the site energy can be obtained from the charge and discharge curves in the OCV measurements, giving us the density of states (DOS): $g(\varepsilon)$. From GITT technique, we obtain the DOS from the relation $dQ/dE_{OCV}$, where the capacity Q (mAh/g) is the total quantity of electrons involved in the electrochemical reaction per unit cell, and $dE_{OCV}$ (V) is determined by the number of sites with energy $\varepsilon$. We calculated the experimental DOS (Fig. 4a) by analyzing the OCV curves (Fig. 3a,b) for the CTF-1, which is formed by multilayer sheets (up to ~10 layers) of 2D polymeric frameworks (Fig. 2c). Then, we studied CTF-1 through first principles calculations on a monolayer and a multilayer system using the all-electron full-potential local orbital (FPLO) code[51, 52], version 9.01-35 within the generalized gradient approximation (GGA).[53] The valence basis set consists of carbon (1s, 2s, 2p, 3s, 3p, 3d), nitrogen (1s, 2s, 2p, 3s, 3p, 3d) and hydrogen (1s, 2s, 2p) states. All atoms including hydrogen atoms were relaxed using force evaluated in a scalar relativistic mode with a convergence criterion of 1 meV/Å. Monolayer is modeled with 20 Å spacing along the *z*-direction, which is large enough to avoid the interaction between the two consecutive interlayers. Self-consistent calculations employed a grid of 6912 (multilayer) and 2500 (monolayer) k points in the full Brillouin zone. From the calculations we extract the theoretical density of states (Fig. 4b) for the 2D (red) and 3D (black) case. Due to the quasi-two dimensional character of CTF-1 multi-layered system, a direct comparison of the experimental band gap, ~1.4 eV, with the monolayer ~2.7 eV, and the multilayer gap, ~1.5 eV, is not relevant. Besides, we can analyze the main features in both DOSs, noticing the existence of a large peak close to the Fermi level, which can indicate the existence of flat bands and charge localization.

One of the most important aspects in the study of CTFs is the ratio of the aromatic rings used in the synthesis process and how the electrochemical properties are affected by this. Then, we synthesized a polymeric framework in which triazine and benzene rings have the ratio of 1:4 (here, we call it as CTF-TCPB. See supporting information Fig. S1 and S2 for the experimental results of CTF-TCPB). We also calculated the DOS and compared the results with the case of CTF-1. From (Fig. S2a in Supporting information) and (Fig. 4b), we observed several peaks in the density of states right below the Fermi level for CTF-1, while in the case of CTF-TCPB, the peaks are shifted to lower energies. This indicates a more conducting behavior for CTF-1 when doping the system (holes) than for that of CTF-TCPB. Indeed, the cyclic voltammetry (CV) measurements for CTF-1 and CTF-TCPB confirmed this result (Supporting information Fig. S2b). While CTF-1 exhibited a continuous CV due to the larger peak-density in the DOS, CTF-TCPB clearly showed two separated redox reactions, due to its different electronic structure. From the latest comparison, we conclude that we can tune the electronic structure properties of CTFs by either changing the periodicity of the frameworks or the aromatic rings in these systems.

In summary, we studied the electronic structure of CTFs, by combining electrochemical experimental techniques and theoretical studies. Our results reveal that the particular electronic structure of CTFs emerges from their structural periodicity and that their electronic properties are controllable by changing the number of layers and the ratio of benzene and triazine aromatic rings. Therefore, covalent-triazine polymeric frameworks can lead to further development of organic semiconductors with tunable electronic properties, where band-gaps and doping properties can be controlled by combining the donor/acceptor character of the starting monomers in the synthesis of these frameworks.

EXPERIMENTAL SECTION

*Synthesis of porous polymeric frameworks.* The CTFs were synthesized by ionothermal synthesis as described in ref. 25; heating a mixture of *p*-dicyanobenzene or tris(4-cyanophenyl benzene), respectively, and $ZnCl_2$ in quartz ampules at 400 °C for 40 hours. The obtained samples were washed with 1 M HCl and distilled water for several times. The nitrogen physisorption measurements were carried out at 77 K up to 1bar using a Quantachrome Autosorb 1C apparatus.

Pore size distribution was obtained by applying the QS-DFT equilibrium model for nitrogen on carbon with slit pores at 77 K.

*Physical characterization for porous polymeric frameworks.* The electron energy-loss spectrum (EELS) measurement was carried out using a Tecnai F30 (FEI company) operated at an accelerating voltage of 300 kV by using a special sample holder which can keep the organic specimen cool. The Raman spectroscopy measurements were carried out by using a NIR-Raman-Spectrometer HoloLab Series 5000 from Kaiser Optical Systems with a laser excitation of 785 nm and a power of 10 mW on the sample.

*Electrochemistry.* The polymeric frameworks were characterized by their electrochemical properties. The electrodes were made by the polymeric framework (70 wt.%), conductive additive (Super-P; 20 wt.%), and binder (carboxyl methyl cellulose; 10 wt.%). We used Al foils as a current collector. We assembled the two-electrode Swagelok-type cells in an argon-filled glove box and tested them on a multichannel potentiostatic-galvanostatic system (VMP-3, Bio-Logic). We used lithium metal as an anode and 1M $LiPF_6$ in ethylene carbonate and dimethyl carbonate (volume ratio 1:1) as an electrolyte.

*Theoretical study.* We performed electronic structure calculations in the framework of density-functional theory (DFT) using full potential local-orbital scheme implemented in FPLO codes[51,52]. The exchange-correlation energy functional was evaluated within the generalized gradient approximation (GGA), using the Perdew, Burke, and Ernzerhof parametrization[53]. We used previous reported data[25] on the crystal structure of CTF-1 and calculated the density of states (DOS) corresponding to the ground state energy of the optimized geometry of the system, for monolayers and bulk multi-layers of CTF-1 and CTF-TCPB compounds. The geometry optimization was carried out by evaluating the atomic forces on each atom with a convergence criterion of 1 meV/Å.

FIGURES

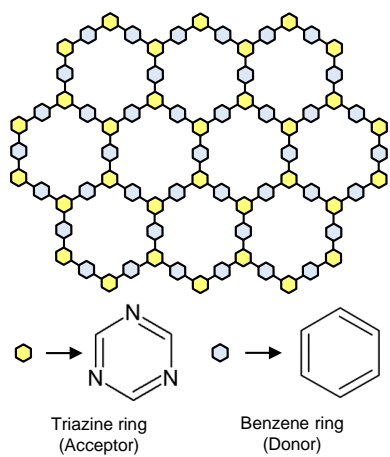

**Figure 1.** Schematic illustration for the ideal CTF constituted by benzene and triazine rings by the ratio of 1:1.5, so-called CTF-1.

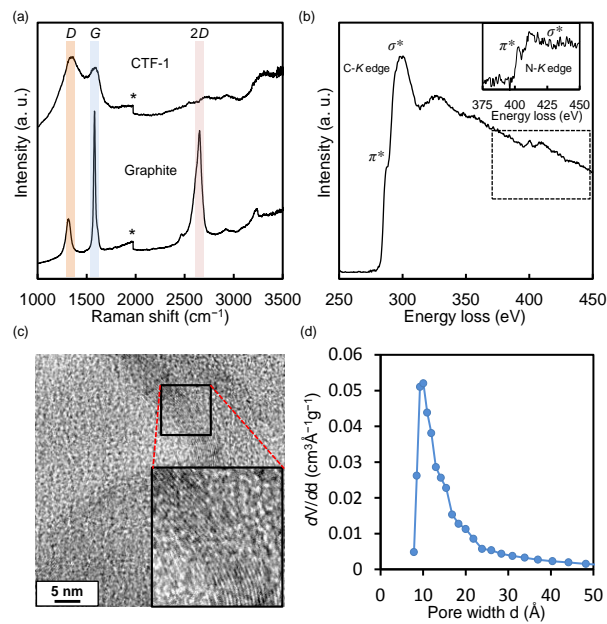

**Figure 2.** (a) Raman spectrum for CTF-1 and graphite, * indicates the artificial step due to the equipment. (b) EEL spectrum for CTF-1. Inset shows the N-*K* edge (the doted area) with the extraction of background. (c) HR-TEM image of CTF-1. Inset shows the edge of CTF-1 showing a stacking of layers. (d) Pore width distribution of CTF-1.

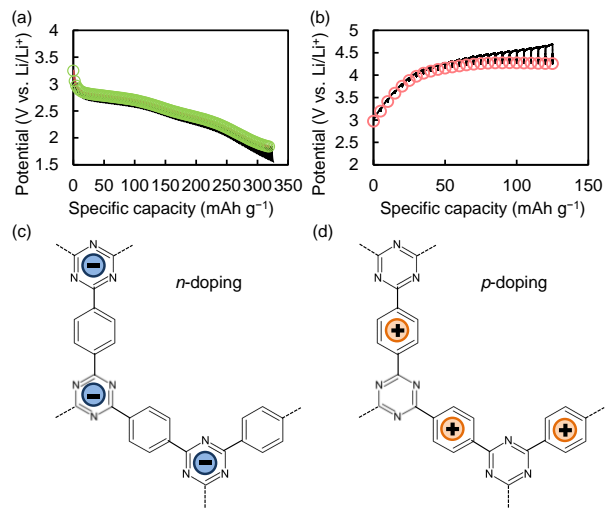

**Figure 3.** (a) OCV curve at *n*-doping. (b) OCV curve at *p*-doping. (c) Schematic illustration of *n*-doping at triazine rings (acceptor). (d) Schematic illustration of *p*-doping at benzene rings (donor).

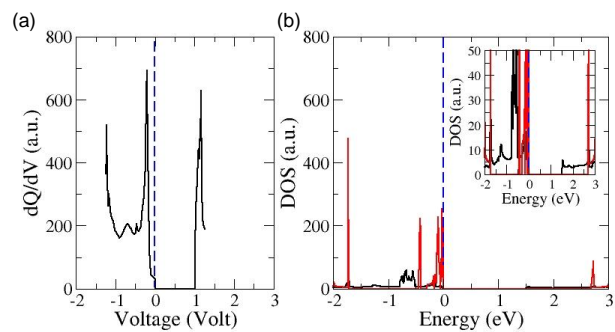

**Fig. 4.** (a) Measurement of the DOS as a function of Voltage for the CTF-1. (b) Theoretical DOS for the ideal CTF-1 for the single layer (red) and the bulk case (black), obtained from *ab initio* calculations. The inset figure presents the same data in a more illustrative fashion. The Voltage (a) and Energy (b) are given respect to the Fermi Level (dashed blue line).

ASSOCIATED CONTENT

**Supporting Information**. Additional figures as described in the text. This material is available free of charge via the Internet at http://pubs.acs.org.

AUTHOR INFORMATION

**Corresponding Author**

*Email: ken.sakaushi@mpikg.mpg.de (K.S.); j.van.den.brink@ifw-dresden.de (J.v.d.B.).

**Notes**

The authors declare no competing financial interests.

ACKNOWLEDGMENT

K.S. would like to thank Dr. Susanne Machill (TU Dresden) for Raman spectroscopy measurements. K.S. is supported by German Academic Exchange Service, DAAD (Grant No.: A/09/74990).

*Supplementary Information*

**Polymeric Frameworks as Organic Semiconductors with Controlled Electronic Properties**

Ken Sakaushi,[†,§,#,*] Georg Nickerl,[§] Hem Chandra Kandpal,[‡,+] Laura Cano-Cortés,[‡]

Thomas Gemming,[†] Jürgen Eckert,[†,¶] Stefan Kaskel,[§] and Jeroen van den Brink[‡,*]

[†]IFW Dresden, Institute for Complex Materials, Helmholtzstr. 20, D-01069 Dresden, Germany

[§]TU Dresden, Department of Inorganic Chemistry, Bergstr. 66, D-01069 Dresden, Germany

[+]Indian Institute of Technology Roorkee, Department of Chemistry, Roorkee-247 667, India

[‡]IFW Dresden, Institute for Theoretical Solid State Physics, Helmholtzstr. 20, D-01069 Dresden, Germany

[¶]TU Dresden, Institute of Materials Science, Helmholtzstr. 7, D-01069 Dresden, Germany

*e-mail: ken.sakaushi@mpikg.mpg.de; j.van.den.brink@ifw-dresden.de

Current address

[#]Max-Planck-Institute for Colloids and Interfaces, Research Campus Potsdam-Golm, D-14424 Potsdam, Germany.

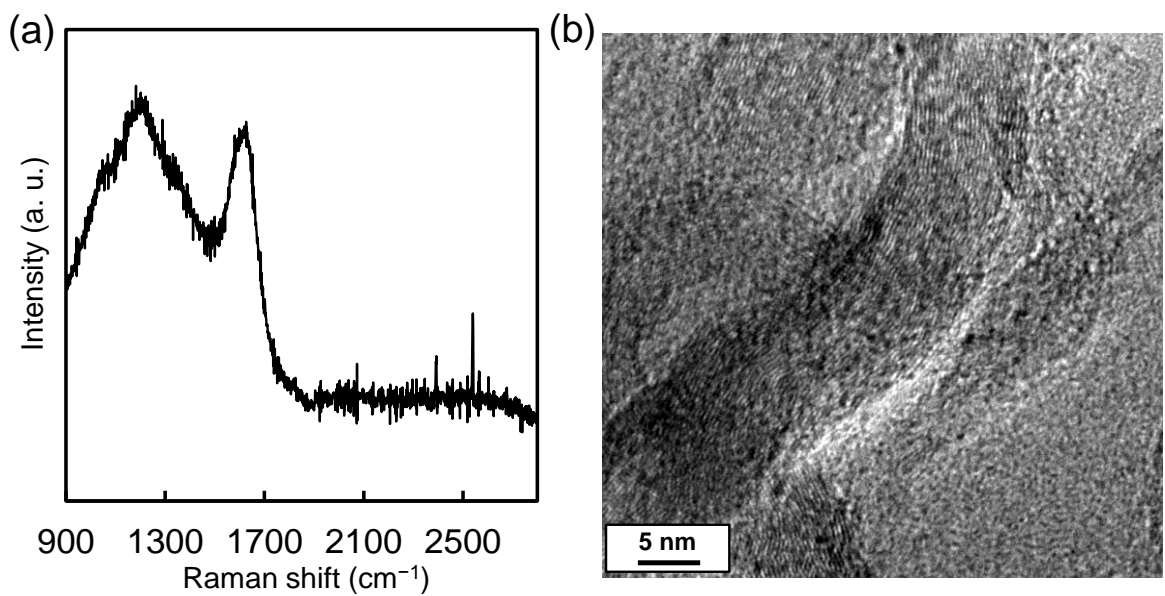

**Figure S1.** (a) Raman spectrum for CTF-TCPB. (b) HR-TEM image of CTF-TCPB.

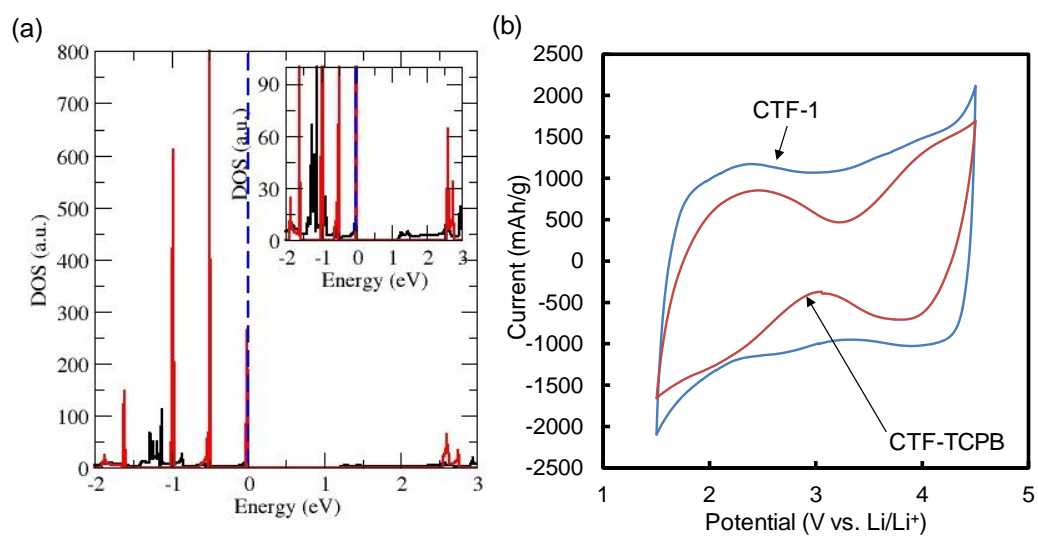

**Figure S2.** (a) Theoretical DOS for the ideal CTF-TCPB for the single layer (red) and the bulk case (black), obtained from *ab initio* calculation. The Voltage and Energy are given respect to the Fermi Level (dashed blue line). b) Cycling voltammogram of CTF-1 and CTF-TCPB.